\documentclass{llncs}
\usepackage{epstopdf}
\usepackage{bm}
  \usepackage{tipa}
  \usepackage[pdftex]{graphicx}
	\usepackage{epsfig}
	\usepackage{epstopdf}
	\usepackage{xcolor}
	\usepackage{amssymb}
\usepackage{tabularx}
\usepackage{graphicx}
\usepackage{multirow}
\usepackage[export]{adjustbox}
\usepackage{float}
\usepackage{algpseudocode,algorithm,algorithmicx}
\usepackage{longtable}
\usepackage{amsmath,amssymb,amsfonts}
\usepackage{graphicx}
\usepackage{epstopdf} 
 
 \usepackage{subcaption}
\usepackage{xcolor}
\usepackage[breaklinks]{hyperref}
\usepackage{url}
\usepackage{breakurl}

\begin{document}

\title{Expanding Cybersecurity Knowledge Through an Indigenous Lens: A First Look}

\author {Farrah Huntinghawk, Candace Richard, Sarah Plosker, Gautam Srivastava
%
\institute{Department of Mathematics and Computer Science, Brandon University, Manitoba, Canada\\ E-mail: \textcolor{blue}{\{huntinfd62,richaric13,ploskers,srivastavag\}@brandonu.ca}\\[1mm]}
}
\maketitle


\section*{Abstract}
 Decolonization and Indigenous education are at the forefront of Canadian content currently in Academia. Over the last few decades, we have seen some major changes in the way in which we share information. In particular, we have moved into an age of electronically-shared  content, and there is an increasing expectation in Canada that this content is both culturally significant and relevant.  In this paper, we discuss an ongoing community engagement initiative with First Nations communities in the Western Manitoba region. The initiative involves knowledge-sharing  activities that focus on the topic of cybersecurity, and are aimed at a public audience. This initial look into our educational project focuses on the conceptual analysis and planning stage. We are developing  a ``Cybersecurity 101'' mini-curriculum, to be implemented over several one-hour long workshops aimed at diverse groups (these public  workshops may include a wide range of participants, from tech-adverse to tech-savvy). Learning assessment tools have been built in to the workshop program. We have created informational and promotional  pamphlets, posters,  lesson plans, and feedback questionnaires which we believe instill relevance and personal connection to this topic, helping to bridge gaps in accessibility for Indigenous communities while striving to build positive, reciprocal relationships. Our methodology is to approach the subject from a community needs and priorities perspective. Activities are therefore being tailored to fit each community.   We hope this will lead to increased awareness and engagement by community members. Two Indigenous student research assistants were hired to assist in this project, which has developed into a blend of community outreach on the topic of security and data protection (most notably with respect to social media and online banking) and a computing education student-led educational research project.

\section{Introduction}
\label{sec:intro}

\thispagestyle{empty}
In 2019, Drs.~Plosker and Srivastava received a Community Investment Grant from the Canadian Internet Registration Authority (CIRA) for a knowledge-sharing community engagement project aimed at broadening public understanding of cybersecurity. The project, which is now underway, is aimed specifically at facilitating broad-interest workshops on cybersecurity in rural First Nations communities in Western Manitoba. Participants of the workshops will gain broad understanding of cybersecurity, as well as practical information on security and data protection. 
The main goal of the project is to include Indigenous peoples in the Canadian cybersecurity dialogue,   couched in the
Four R's of First Nations and Higher Education: Respect, Relevance, Reciprocity, and Responsibility \cite{4R}.

There are 63 First Nations in the province of Manitoba. In 2016, census data shows that Manitoba had a population of 1,240,695 people, with 223,310 	or 18.0\% identifying as Aboriginal \cite{census}. The term ``Western Manitoba" is somewhat loosely defined, but comprises Treaties 1, 2, 4, and 5. Brandon University is located on Treaty 2 Lands,  traditional homelands of the  Anishinabek, Cree, Dakota, Dene, Oji-Cree,  and M\'etis peoples.

Cybersecurity in Manitoba has made headlines recently: At the time of writing, the province of Manitoba has announced a commitment of over half a million dollars to help create a Cyber Security Technical Centre of Excellence \cite{CBC3}. In the news release, Economic Development and Training Minister Ralph Eichler states ``Cyberattacks are on the rise, and experts predict that cybercrime damages will cost the world more than \$6 trillion annually by 2021''. Other recent notable cybersecurity news in Manitoba include a ransomware attack on an insurance and financial brokerage based in Manitoba, that did not report the attack to the Office of the Privacy Commissioner \cite{CBC4}, and a cyberattack on the overseeing body of foster children in ten agencies in southern Manitoba, made possible because specific information on their IT system and software was published publicly on their website \cite{CBC5}.  

Our project is incredibly timely: with a planned  \$83.9 million infrastructure investment  to provide a fibre-optic cable network for internet service into 112 rural Manitoba communities, including 48 First Nations \cite{CBC1,CBC2}, the communities we visit will soon have faster, more reliable internet. It is therefore   beneficial to discuss email phishing, password security, and other important and relevant aspects of cybersecurity preemptively.

This project is in its initial stages. Although the project is still underway, we describe here in detail our work done thus far, as we feel that knowledge transfer of this project will have a positive impact
on the larger Computer Science Education sphere within Canada. We note that it is important to understand how much time and effort is put into the organizing, planning, and preparing stage, so that educational activities are ultimately successful. Aligning educational resources in
cybersecurity  with the Truth and Reconciliation Commission's Calls to Action, and disseminating these resources and initial findings, will  benefit  
those involved in the broader Canadian Education system.

\section{Project Methodology and Description}
Drs.~Plosker and Srivastava are both Associate Professors in the Department of Mathematics and Computer Science at Brandon University in Brandon, Manitoba, Canada. Dr.~Plosker has a research background in applied mathematics, including quantum cryptography. Dr.~Srivastava has a research background in cybersecurity. Both are interested in Indigenous community outreach.

The initial grant application for this project involved consultations with Chris Lagimodiere, Director of the Indigenous People' Center at Brandon University, to help ensure Indigenous cultures and communities are treated respectfully and to ensure a focus on community needs. We decided to limit our project to First Nations communities that are within a two hour drive from Brandon, Manitoba, so that we could reasonably drive to the communities to facilitate on-site workshops. We feel that it is important to go to the communities rather than invite community members to come to Brandon to attend workshops, in part from a logistics standpoint (to increase participation) and from a standpoint of respect and consideration.

Two Brandon University students were hired to assist with the project as Indigenous outreach workshop facilitators. One of the major candidate  qualifications was knowledge of Indigenous communities in the Western Manitoba area. Huntinghawk (of Rolling River First Nations) and Richard (of Sandy Bay First Nation) were hired. Huntinghawk had previously worked under Plosker in facilitating a series of Fun Math Workshops at the Indigenous Peoples' Center on campus in 2016-17 (e.g.\ Sudoku tournament, $\pi$-related activities for March 14, etc). Though it was not mandatory in terms of the job description and candidate qualifications, both of the student assistants employed for this project are Indigenous. 

The student assistants have worked closely on the creation and Indigenization of the handouts and other educational materials related to the project in a respectful manner. These educational materials used for the workshops can be found in the supplemental material at \cite{bu} and contain basic information on cybersecurity for a public audience. The educational materials have gone through a series of editorial iterations; as we facilitate the cybersecurity workshops, we notice what works well and what does not, and have made changes based on our observations. We have, to the best of our abilities, aligned the educational resource materials with the Truth and Reconciliation Commission's Calls to Action \cite{TRC}, specifically with regard to Education: to develop culturally appropriate curricula. The pamphlet was written to be both entertaining and informative. Rather than being a sterile list of do's and don't's, the pamphlet reads more like a friendly chat: ``Canadians love social media, eh?'' starts off the section on social media security. The pamphlet mentions the usefulness of social media in terms of following the goings-on of comedians Don Burnstick, Wabigut and other Native icons, before cautioning the reader to avoid posting too much personal information on social media, and recommending having a strong password. The pamphlet also includes the logo of the reserve that we visit, and can be further customized as needed.

At the time of writing, we have completed the initial stages of the project.  We are working closely with the Brandon University Research in Ethics Committee (BUREC) and have received ethics certification for the project. Plosker and Srivastava received certificates of completion of the Tri-Council Policy Statement: Ethical Conduct for Research Involving Humans (TCPS) Course on Research Ethics (CORE) tutorial. 
We have had positive initial discussions and interactions   with band Chiefs,  Council members, Directors of Education, school principals, and other designates. We have created survey questionnaires (with feedback from BUREC), educational pamphlets, promotional posters, and lesson plans. The actual ``Cybersecurity 101'' content can be customized based on whether we provide a workshop in a school, for youth learning about electrical and computer engineering, or a community-wide workshop, aimed more at adults and seniors. All educational  resource materials created have some flexibility built in, so they can be edited based on comments from the band Chief, Council, school principal, instructor, Elders, or others, and can be tweaked on an as-needed basis. 

We have, at the time of submitting this article, facilitated workshops within  the community of Waywayseecappo. A workshop within  Sandy Bay First Nation had to be rescheduled due to circumstances outside of our control. We are in the process of finalizing dates for the community of Sandy Bay as well as the remaining workshops.

\section{Related Work}

There is very little literature on the subject of Indigenous education initiatives in the area of electrical and computer engineering, let alone cybersecurity specifically. Much of the computer science literature surrounding Indigenous cultures relates to preserving their languages through digital technologies, or Indigenous health. Alternatively there are examples of integrating Indigenous culture with digital technology. A community in Alaska is  respectfully integrating a part of Alaskian Indigenous culture by combining digital story-telling with cancer education  \cite{Storytelling}.
    
Numerous articles exist on the topic of cybersecurity and social media as it relates to Indigenous youth in Australia: a recent literature review of 22 articles related to  `social media and/or digital technologies and Indigenous Australians'' \cite{22} found several major themes surrounding ``how and why Indigenous young people use social media: identity, power and control, cultural compatibility and community and family connections''. Positive applications of social media and digital technology included health promotion and marketing, whereas negative applications included cyberbullying, cyber racism, and the exchange of sexually explicit content between minors.

With regard to Canadian Indigenous peoples, research includes interviews of Iqaluit students showing how their use of social media---in particular, Facebook---was closely linked to their identity \cite{IqaluitFB}. Similarly, Inuit young adults are using Youtube as a digital storytelling medium \cite{InuitYT}.  An in-depth literature review supplemented by personal communications on the use of digital technologies in remote    and    northern    Indigenous    communities    in    Canada \cite{Northern}  analyses uses and challenges of technology in indigenous communities. The paper develops a community technology adoption assessment and identifies the different levels of technology adoption throughout indigenous communities.

Using five First Nations communities as case studies, four in Ontario and one in British Columbia,   Polyzois asked ``How are First Nations communities using
online tools to enable   residents to more actively participate in the planning of their community?" \cite{Poly}.  The work includes a thorough literature review of over 300 documents, through which two central themes emerged: public engagement and Indigenous planning. In relation to these two themes, a number of the documents discuss  the \emph{digital divide} between
First Nations and other communities in Canada, defined in \cite{Poly} as  social and 
economic   inequality  with respect to access to and use of information and computer
technologies (ICT). The study found a number of challenges: digital connectivity, internet accessibility, affordability, and 
quality of ICT service. It also identified a number of opportunities: ownership and control of ICT, building capacity and
developing technical skills, cultivating social capital, and
fostering community resilience. As noted in \cite{Poly}, addressing some of these challenges and opportunities would be ``a key step in moving closer to establishing new best practices in First Nation
community development in rural and remote locations in Canada''. 

The work of \cite{McM} teases out the link between ``networked digital infrastructure development and the autonomy and agency of Indigenous peoples''. It should be noted that the majority of the references mentioned herein are work done not within Computer Science departments, but within the  areas of Health, Geography, Sociology, and Communications.  

\subsection{Data Collection}
We have found it easiest to start the workshops in communities where one of the two student Indigenous outreach workshop facilitators (Huntinghawk and Richard) have personal connections: Waywayseecappo and Sandy Bay. For these two communities, Richard has worked closely with educational representatives on finalizing dates and other logistics.  The first workshops were at Waywayseecappo Community School, one for both Grade 7 classes, and one for both Grade 8 classes. The next workshop will be  at  Isaac Beaulieu Memorial School in Sandy Bay, aimed at Grades 11 and 12. These grades were chosen by the representatives that our team was in contact with in the respective communities.

 Waywayseecappo First Nation was established through the signing of Treaty 4 in 1874. Situated near the southwest corner of Riding Mountain National Park, it is within 8 km of the town of Rossburn, and consists of 24,600 acres. The total registered (status) population living in Waywayseecappo is 1,365 (2016 data) \cite{WWStatCan}.

  The Sandy Bay Ojibway First Nation was established through the signing of Treaty 1 in 1871, with about 180 people of mixed-blood ancestry. Sandy Bay is situated on Reserve No.~5, which comprises 16,456 acres on the western shore of Lake Manitoba. The total registered population of Sandy Bay is 6,174 (July, 2013 data) and is increasing. See \cite{SandyBay} for further details.

We plan to provide workshops in several other First Nations communities surrounding Brandon and potentially at local institutions. As a first step, we have contacted and have had positive response from several communities that have pre-established relationships with Brandon University (through the Student Recruitment office).   Our team is working to finalize dates of the remaining workshops. 

We have constructed pre- and post-workshop survey questionnaires for workshop participants; the former to gather information so as to learn how much the community members already know about cybersecurity and how important they think it is, and the latter to test how effective our workshops are, so that we can make changes as necessary as we go forward with the project. 
Questions range from basic administrative data (age, sex, name of community) to questions about smartphone, social media, and internet use, whether they sign out of online accounts or just close the web browser, how important they feel internet safety is, and whether or not they've experienced cyberbullying, among other things (a full list of survey questions is available at \cite{bu}).  The draft questionnaires have been given to our community contacts in advance, for feedback and customization (e.g.\ if a particular community is having issues surrounding social media, more questions related to that topic could be added to the questionnaire in advance of the workshop).    

Consent forms to use the data have been developed and are sent to the instructor or designate prior to the workshop, for signature by the guardians. 
To complement the workshops, we are also hoping to Indigenize some of the ``Cybersecurity 101'' materials which we are using at the workshops and that could potentially be used in introductory computer science courses at Brandon University as well.  

Most research in cybersecurity education uses analogies: sealed envelopes, boxes locked with padlocks,
malicious eavesdroppers/adversaries, colour-coding, and the use of signatures, seals, or fingerprints. This
education framework does not take Indigenous knowledge-sharing and storytelling into account. We plan to work closely with several Elders and Knowledge-Keepers in the area to Indigenize the cybersecurity   educational materials  through the development of
cybersecurity educational analogies using culturally relevant stories and ideas.

\subsection{Strengths \& Limitation}
Awareness, education, and training can minimize or prevent cybersecurity risks. Ensuring that the public understands the importance of being vigilant is key to obtaining ``buy-in'' to attending and participating in educational activities. For this to happen, the topics and ways of approaching the topics must be made personally relevant. For our project, the context so far has been  Treaty 1 and Treaty 4   First Nations communities that will be receiving new fibre optics infrastructure for high-speed internet. Brandon University's main campus is located on Treaty 2 land, with roughly 13\% of our students self-identifying as Aboriginal (First Nations, M\'etis, or Inuit)\cite{IPC}, and has a general positive connection with many Indigenous communities in the Western Manitoba region (e.g., our recruitment officers regularly recruit students living on First Nations reserves, there is a weekly community traditional beadwork event hosted at the Indigenous Peoples' Center on campus, a regular graduation Pow Wow to honour Indigenous graduates, etc).  For more information on our Indigenous Peoples' Center and its services, see \cite{IPC}. 

Individuals or groups looking to implement similar initiatives will need to reflect on the realities of their geographic location, the relationship their institution has developed with Indigenous communities in the area, and the needs of these communities. 
Researchers and educators will find that one cannot take our educational materials and ``skip to the finish line''; it is a process. Time needs to be invested into creating meaningful, reciprocal relationships with community members.  All materials should be tailored to each community.

\section{Conclusions and Future Work}
For this project, we felt it imperative that our work was guided by an Indigenous knowledge framework. The creation of new educational materials on cybersecurity is being done so as to incorporate culturally appropriate and relevant Indigenous knowledge-sharing in a respectful way. We believe this will lead to increased community engagement and level of satisfaction with the cybersecurity activities. The educational materials were first written in skeletal form, so that they could be edited and tailored to fit each community and each situation. 

We found it important to emphasize relationship building during this research project. Researchers should not be motivated solely to obtain results; it is about the entire process. Having pre-established relationships through students  involved in the project, or through various offices at one's institution, helps but we caution that it does not instantaneously guarantee success.

At present, we have funding to visit five First Nations communities in Western Manitoba. We hope to secure more funding for future events, both for visiting more communities as well as returning to communities to ensure a continued, lasting relationship.

Upon completion of the workshops, we plan to reflect on not only the results of the survey questionnaires but also the  process as a whole.
Workshop lesson plans and educational materials created through this initiative have been made freely available at \cite{bu}. We encourage any interested groups to tailor the activities and handouts to their geographic region. 

\section{Acknowledgements}
This work was supported by a Community Investment Grant from the Canadian Internet Registration Authority (CIRA). The authors thank Chris Lagimodiere for discussions at the initial stage of this project, as well as the Brandon University Research Ethics Committee (BUREC) for guidance. C.~Richard was supported by a travel grant through the Sandy Bay Band office.

\bibliographystyle{splncs03}
\bibliography{ref}

\end{document}